\documentclass[runningheads]{llncs}

\usepackage{accv}

\usepackage{accvabbrv}

\usepackage{graphicx}
\usepackage{booktabs}

\usepackage[accsupp]{axessibility}  % Improves PDF readability for those with disabilities.

\usepackage{hyperref}

\usepackage{orcidlink}

\begin{document}

\title{Window-based Channel Attention for Wavelet-enhanced Learned Image Compression} 

\titlerunning{Window-based Channel Attention for Wavelet-enhanced LIC}

\author{
Heng Xu\inst{1}\orcidlink{0000-0001-6027-6365} \and
Bowen Hai\inst{1}\orcidlink{0009-0007-7130-107X} \and
Yushun Tang\inst{1}\orcidlink{0000-0002-8350-7637} \and
Zhihai He\inst{1,2*}\orcidlink{0000-0002-2647-8286}
}


\institute{Department of Electronic and Electrical Engineering, Southern University of Science and Technology, Shenzhen, China \and
Pengcheng Laboratory, Shenzhen, China\\
\email{xuh2022@mail.sustech.edu.cn, haibw2022@mail.sustech.edu.cn, tangys2022@mail.sustech.edu.cn, hezh@sustech.edu.cn}
}
\renewcommand{\thefootnote}{\fnsymbol{footnote}}
\footnotetext[1]{~Corresponding author.}
\renewcommand{\thefootnote}{\arabic{footnote}}
\maketitle

\begin{abstract}
  Learned Image Compression (LIC) models have achieved superior rate-distortion performance than traditional codecs. Existing LIC models use CNN, Transformer, or Mixed CNN-Transformer as basic blocks. However, limited by the shifted window attention, Swin-Transformer-based LIC exhibits a restricted growth of receptive fields, affecting the ability to model large objects for image compression. To address this issue and improve the performance, we incorporate window partition into channel attention for the first time to obtain large receptive fields and capture more global information. Since channel attention hinders local information learning, it is important to extend existing attention mechanisms in Transformer codecs to the space-channel attention to establish multiple receptive fields, being able to capture global correlations with large receptive fields while maintaining detailed characterization of local correlations with small receptive fields. We also incorporate the discrete wavelet transform into our Spatial-Channel Hybrid (SCH) framework for efficient frequency-dependent down-sampling and further enlarging receptive fields. Experiment results demonstrate that our method achieves state-of-the-art performances, reducing BD-rate by 18.54\%, 23.98\%, 22.33\%, and 24.71\% on four standard datasets compared to VTM-23.1.
  \keywords{Learned Image Compression \and Window-based Channel Attention \and Receptive Field \and Wavelet Transform}
\end{abstract}

\section{Introduction}
\label{sec:intro}
As the resolution of digital images continues to increase, image compression techniques play a crucial role in computer storage and transmission.
Traditional image compression codecs such as JPEG2000 \cite{taubman2002jpeg2000}, BPG \cite{bellard2014bpg}, and VVC \cite{wien2020versatile} are now widely utilized. 
In recent years, Learned image compression (LIC) \cite{balle2016end,ballé2018variational,minnen2018joint,minnen2020channel,cheng2020learned,pan2022content,he2021checkerboard,xie2021enhanced} models have shown significant advancements, surpassing existing traditional codecs in terms of various metrics. 
This progress suggests that LIC may emerge as the next-generation technology of image compression.

Convolutional neural networks (CNN) have demonstrated competence in LIC. 
Ballé \etal \cite{balle2016end} built a basic framework for CNN-based LIC, which was further extended by incorporating VAE and hyper-prior modules to enhance the performance \cite{ballé2018variational}. 
To improve the entropy coding performance, auto-regressive models \cite{minnen2018joint} and Gaussian mixture models (GMM) \cite{cheng2020learned} have been proposed. 
Transformer-based models, leveraging the advantages of Transformer architectures in computer vision tasks, have emerged as promising alternatives in LIC \cite{zhu2021transformer,qian2021entroformer,zou2022devil,koyuncu2022contextformer,liu2023learned}. 
These models, such as Swin-Transformer-based model \cite{zhu2021transformer} and parallel bidirectional context Transformer-based model \cite{qian2021entroformer}, have demonstrated superior compression performance over classic CNN-based approaches. 

Numerous studies have sought to enlarge receptive fields of their CNN-based models to improve the performance for computer vision tasks \cite{tsai2018learning,fu2018deep,singh2018analysis,plotz2018neural,shi2020pv}. 
For recent Transformer models, Xie \etal \cite{xie2021segformer} noted that their MLP decoder benefits from Transformers having a larger effective receptive field than other CNN models, corresponding to the performance gain of Transformers in LIC.
However, Xia \etal \cite{xia2022vision} argued that Swin-Transformer model \cite{liu2021swin} exhibits a restricted growth of receptive fields due to the shifted window attention mechanism, limiting its capability to model large objects effectively. 
Meanwhile, LIC models encode and decode the entire image containing both small objects and large objects, requiring large receptive fields to improve the performance.
% However, constrained by the window size in Swin-Transformer, the receptive field of the Transformer model is only slightly larger than that of CNN. This limitation poses a challenge in capturing long-range dependencies in the image. As a result, the models may struggle to effectively focus on complex spatial relationships across the entire image, particularly in LIC where comprehensive understanding of the image content is crucial for achieving low bit rates and high image quality. The restricted receptive field can lead to suboptimal performance and may hinder the Transformer's ability to compete with CNN-based approaches in image compression tasks. 

\begin{figure}[tb]
  \centering
  \begin{subfigure}{0.49\linewidth}
  \centering
    \includegraphics[width=0.95\textwidth]{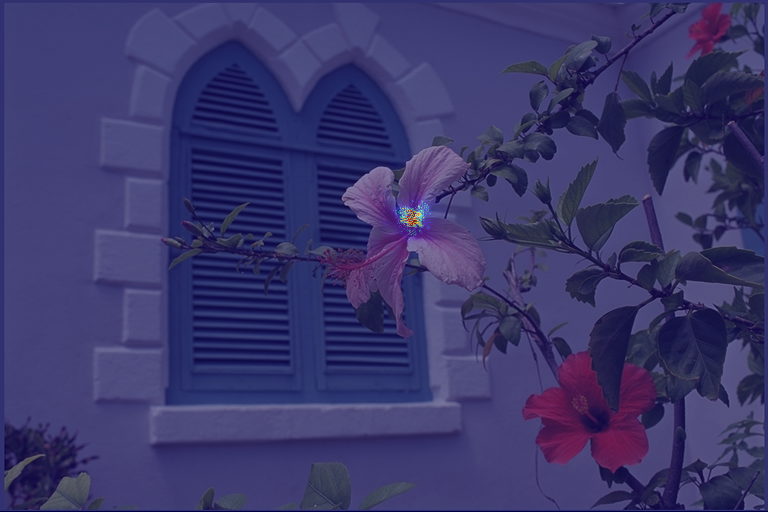}
    \caption{Residual block from \cite{he2016deep}}
    \label{fig:erf_a}
  \end{subfigure}
  \hfill
  \begin{subfigure}{0.49\linewidth}
  \centering
    \includegraphics[width=0.95\textwidth]{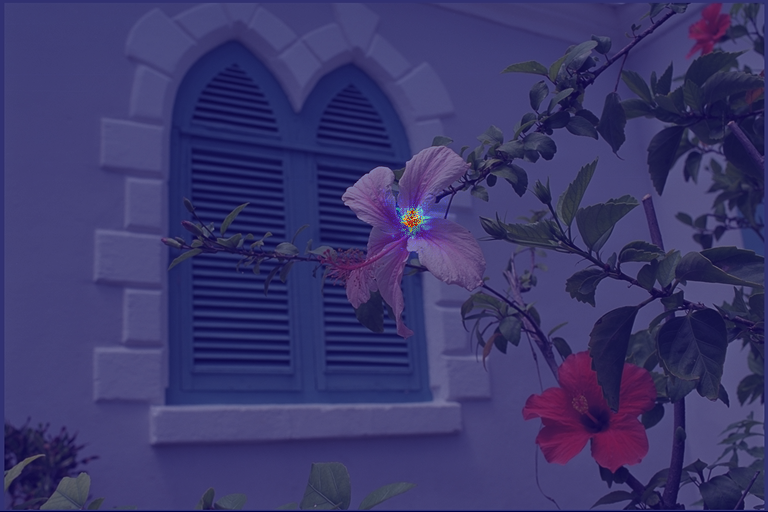}
    \caption{Space attention from \cite{liu2021swin}}
    \label{fig:erf_b}
  \end{subfigure}\\
  \begin{subfigure}{0.49\linewidth}
  \centering
    \includegraphics[width=0.95\textwidth]{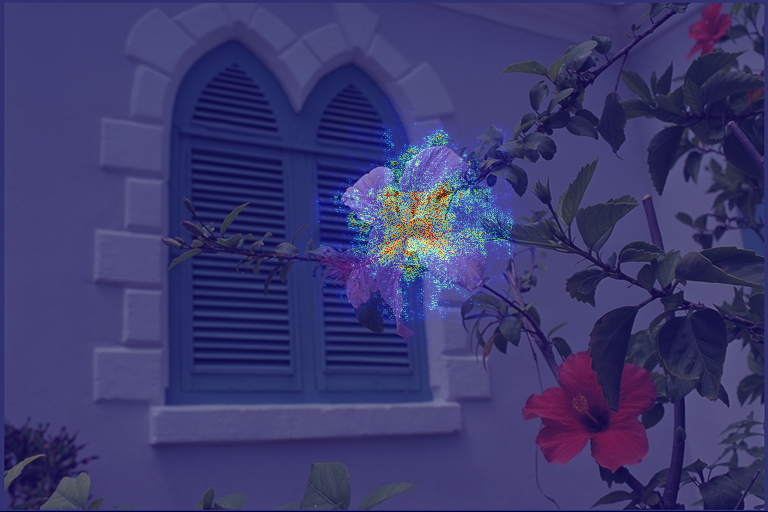}
    \caption{Ours}
    \label{fig:erf_c}
  \end{subfigure}
  \hfill
  \begin{subfigure}{0.49\linewidth}
  \centering
    \includegraphics[width=0.95\textwidth]{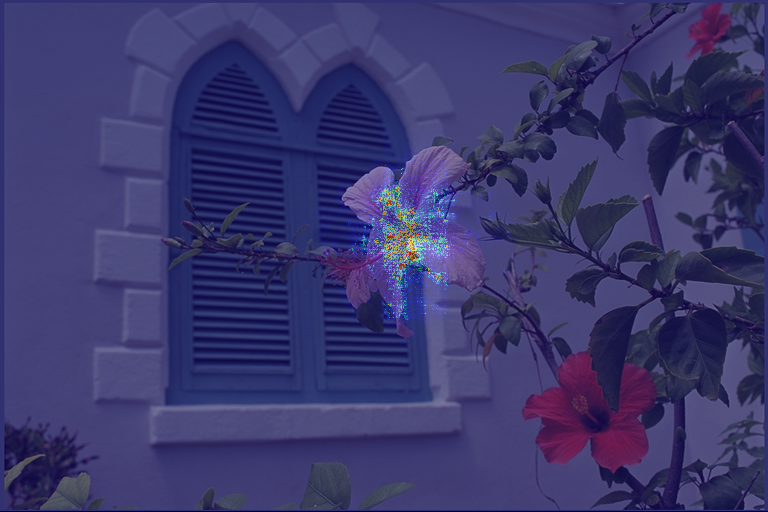}
    \caption{Ours without wavelet transform}
    \label{fig:erf_d}
  \end{subfigure}
  \caption{Effective Receptive Fields (ERF) on \emph{kodim07} from modules of our SCH block. (a) and (b) are from \cite{he2016deep,liu2021swin}, while (c) and (d) are our window-based channel attention with and without wavelet transform. Results are normalized and clipped by a threshold of 0.3 for better visualization. The color changes from blue to red as the value increases.}
  \label{fig:erf}
\end{figure}

In this paper, we initially integrate the window attention of Swin-Transformer with CNN to better capture local dependencies for small objects in the image. 
To address the issue of limited receptive fields and improve the performance, we explore the attention across the channel dimension, given that channel tokens inherently contain global spatial information, as shown in Fig.\ref{fig:SCdemo}. 
Specifically, we replace the shifted window attention of Swin-Transformer with the proposed window-based channel attention, aiming to capture more global dependencies than other channel attention methods and enlarge receptive fields, as visualized in Fig.\ref{fig:erf}.
However, Ding \etal \cite{ding2022davit} stressed that global channel tokens impede local interactions across spatial locations. 
Hence, we need to incorporate channel attention with local modules to learn both local and global information, corresponding to the combination of residual blocks \cite{he2016deep}, space attention modules \cite{liu2021swin}, and channel attention modules in our Space-Channel Hybrid (SCH) block.  

Other works introduced Discrete Wavelet Transform (DWT) to effectively enlarge receptive fields for computer vision tasks \cite{liu2018multi,liu2019multi,wang2020multi,jeevan2022wavemix,yao2022wave,liu2023learned}. 
DWT decomposes the image into sub-images with different frequency properties, functioning as a parameter-free down-sampling method to enlarge receptive fields. 
DWT has also been explored in \cite{ma2019iwave,ma2020end,sahin2023image} to efficiently encode images and optimize bit allocation for image details in LIC.
In this work, we propose to use DWT to further enlarge receptive fields for our window-based channel attention as shown in Figs.\ref{fig:erf_c} and \ref{fig:erf_d}, thereby improving the performance of our LIC system.

The contributions of our work are summarized as follows:
\begin{itemize}
\item We propose a novel Space-Channel Hybrid (SCH) framework for LIC, containing residual blocks and space attention modules for local information learning, and channel attention modules for global information learning.
\item We are the first to incorporate window partition into channel attention, aiming to capture more global information and effectively enhance receptive fields for LIC. To further enlarge receptive fields, we integrate the DWT module with the residual block for our framework.
\item With the above techniques, our method achieves state-of-the-art performances across four datasets with various resolutions. Compared to the anchor VVC (VTM-23.1) \cite{wien2020versatile}, our method provides a reduction in Bjøntegaard-delta-rate (BD-rate) \cite{bjontegaard2001calculation} by 18.54\%, 23.98\%, 22.33\%, and 24.71\% on Kodak \cite{Kodak1992}, Tecnick \cite{asuni2014testimages}, CLIC Pro Val \cite{toderici2020workshop}, and CLIC 2021 Test \cite{toderici2021workshop}.
\end{itemize}

\section{Related Work}
\subsection{Learned Image Compression}
\subsubsection{CNN-based Models}
In recent years, there have been notable breakthroughs in CNN-based LIC \cite{balle2016end,ballé2018variational,minnen2018joint,Nakanishi2019,minnen2020channel,cheng2020learned,pan2022content,he2021checkerboard,xie2021enhanced}.
Ballé \etal \cite{balle2016end} built a basic CNN-based framework for end-to-end LIC models. 
They extended the work by introducing a VAE architecture and incorporating a hyper-prior module to enhance the capabilities of LIC \cite{ballé2018variational}.
Drawing inspiration from the success of auto-regressive priors, Minnen \etal \cite{minnen2018joint} introduced an auto-regressive component to improve entropy modeling.
Cheng \etal \cite{cheng2020learned} substituted the Single Gaussian Model (SGM) in the entropy model with a Gaussian Mixture Model (GMM), which involved the integration of residual blocks and a simplified attention module.
Addressing the computational overhead of the context model, He \etal \cite{he2021checkerboard} introduced a checkerboard context model designed for parallel computing. 
Minnen \etal \cite{minnen2020channel} reduced computational costs by employing a channel-wise context. 
Beyond advancements in entropy modeling, researchers have explored diverse CNN architectures to elevate feature extraction for LIC. 
Liu \etal \cite{liu2019non} introduced non-local residual blocks to capture local and global correlations, leveraging attention masks to allocate bits intelligently based on feature importance. 
Xie \etal \cite{xie2021enhanced} exploited invertible neural networks (INNs) to enhance overall performance.
Pan \etal \cite{pan2022content} introduced the content adaptive channel dropping (CACD), aiming to improve the content adaptability on both latents and the decoder.

\subsubsection{Transformer-based Models}
With the continuous evolution of the Transformer architecture, its advantages in computer vision have been progressively explored and applied in tasks such as image classification \cite{chen2021crossvit}, object detection \cite{carion2020end}, image reconstruction \cite{cai2022coarse}, etc. 
Similarly, Transformer has also found applications in LIC tasks \cite{zhu2021transformer,qian2021entroformer,zou2022devil,koyuncu2022contextformer,liu2023learned}.
Zhu \etal \cite{zhu2021transformer} proposed Swin-Transformer-based image compression, demonstrating their superiority with convolutional counterparts. 
Qian \etal \cite{qian2021entroformer} introduced a parallel bidirectional Transformer-based context model while maintaining time efficiency. 
Zou \etal \cite{zou2022devil} proposed a window-based attention mechanism to enhance Transformer-based LIC models.
Koyuncua \etal \cite{koyuncu2022contextformer} proposed a Transformer-based context model that utilizes multi-head attention to model the entropy in the latent space, allowing for adaptive entropy modeling of spatial and cross-channel dependencies. 
Liu \etal \cite{liu2023learned} proposed an efficient parallel Transformer-CNN Mixture (TCM) block with a controllable complexity to incorporate the local modeling ability of CNN and the non-local modeling ability of Transformer to improve the performance.
To combine the advantages of CNN and Transformer, we incorporate residual blocks and space attention modules into our SCH framework to better capture local dependencies in image features.

% \subsection{Attention Mechanism}
% \subsubsection{Space Attention}
% In the realm of LIC, space attention mechanisms play a pivotal role in guiding models toward salient regions, thereby aiding in the extraction of finer details. Several attention modules tailored for image compression have demonstrated significant enhancements in rate-distortion (RD) performance.
% Liu \etal \cite{liu2019non} introduced non-local operations aimed at capturing both local and global correlations, leveraging attention masks to allocate bits intelligently based on feature importance. Inspired by Swin-Transformer, Zou \etal \cite{zou2022devil} proposed window-based attention, which combines local-aware attention mechanisms with global-related feature learning.
% Moreover, Liu \etal \cite{liu2023learned} combined the CNN and Swin-Transformer, offering a potentially more effective approach for feature fusion.

\subsection{Channel Attention}
Channel attention primarily concentrates on the correlations among feature channels, modeling the significance of each channel globally across the space dimension. 
This mechanism is mainly utilized in various computer vision tasks. 
Hu \etal \cite{hu2018squeeze} introduced Squeeze-and-Excitation Networks (SENet), aiming to dynamically learn the importance of individual channels through channel attention mechanisms, which are based on global average pooling and CNN.
Woo \etal \cite{woo2018cbam} extended the capabilities of SENet by adding the max-pooling operation, enhancing the overall representation of features.
Ding \etal \cite{ding2022davit} proposed a Transformer-based channel group attention mechanism that complements its space window attention by offering a global receptive field on the whole-image size. 
% This augmentation enables the model to capture extensive visual features and extract global information while concurrently preserving the capability to model local dependencies.
Existing works squeezed the spatial dimension of the entire feature map to obtain the channel attention map, but we argue that it leads to excessive information loss and limits the variety of channel information expression.
In this work, we introduce window partition to channel attention module to learn more diverse global information. By computing different channel attention maps for different windows, this module is superior to existing channel attention modules.

\subsection{Wavelet Transform for Deep Neural Networks}
For LIC, Ma \etal \cite{ma2019iwave} introduced a model that employed a trained CNN as a filter to emulate a wavelet-like transform. 
Ma \etal \cite{ma2020end} further refined their approach, using the wavelet-like transform to convert images into coefficients without information loss, which can be optionally quantized and encoded into bits for image compression. 
For computer vision tasks, many works utilized Haar wavelet to enlarge receptive fields because of its simplicity and efficiency \cite{liu2018multi,liu2019multi,wang2020multi,jeevan2022wavemix,yao2022wave,li2024ewt}. 
Liu \etal \cite{liu2018multi} adopted Haar DWT as default to effectively enlarge receptive fields without information loss for image restoration. 
Jeevan \etal \cite{jeevan2022wavemix} concluded that Haar DWT assists the model in enlarging receptive fields faster than convolutional down-sampling.
Li \etal \cite{li2024ewt} utilized Haar DWT and IDWT for effective downsampling and upsampling, which benefits image restoration because of enlarged receptive fields and frequency-related information learning.
In this work, we propose to explore Haar DWT to effectively enlarge receptive fields for more efficient image compression.

\section{Method}
\subsection{Problem Formulation}
Our model is built upon the channel-wise auto-regressive entropy model \cite{minnen2020channel} as shown in Fig.\ref{fig:arch}.
We first introduce the workflow of the basic LIC.

Given an input image $x$, the analysis transform $g_a$ maps it to a latent representation $y$, which is then quantized into $\lceil y \rfloor$ by $Q$, and we employ a range coder to encode it losslessly. 
% Ballé \etal \cite{ballé2018variational} designed the factorized entropy model to estimate the probability distribution of each symbol in $\lceil y \rfloor$. 
According to \cite{minnen2018joint}, we encode $\lceil y-\mu \rfloor$ instead of $\lceil y \rfloor$, where $\mu$ is the mean estimated by the entropy model. 
Suppose $y_q=\lceil y-\mu \rfloor$, we then reconstruct the coded $\hat{y}$ as $y_q + \mu$, and it is mapped back to the reconstructed image $\hat{x}$ using the synthesis transform $g_s$. 
The main pipeline is formulated as:
\begin{align}
  y = g_a(x;\phi), \quad y_q = Q(y-\mu)&, \quad \hat{y} = y_q + \mu, \quad \hat{x} = g_s(\hat{y}; \theta),
\end{align}
where $g_a$ and $g_s$ are parameterized by $\phi$ and $\theta$, respectively. 
The latent residual prediction $r$ \cite{minnen2020channel} is added to compensate for quantization errors.

% To better estimate the probability, Ballé \etal \cite{ballé2018variational} proposed the hyper-prior path to compute the hyper-prior $z$, which is next quantized and entropy coded as side information. 
% Each element in $\hat{y}$ is modeled as an independent Gaussian distribution, with mean and variance computed from the quantized $\hat{z}$. 
% Therefore, the probability of the $i$th element is $p(\hat{y}_i|\hat{z}_i)=\mathcal{N}(\mu_i,\sigma_i^2)$ so that spatial redundancy is reduced in probability estimation. 
% The hyper-prior pipeline is formulated as:
% \begin{align}
%   z &= h_a(y;\phi_h) \\
%   \hat{z} &= Q(z) \\
%   (\mu, \sigma) &= h_s(\hat{z}; \theta_h)
% \end{align},
% where the hyper-prior analysis transform $h_a$ and the hyper-prior synthesis transform $h_s$ of hyper-prior are parameterized by $\phi_h$ and $\theta_h$, respectively. 

Ballé \etal \cite{ballé2018variational} proposed the hyper-prior path to compute the hyper-prior $z$ as side information. 
The probability of each element in $\hat{y}$ is modeled as an independent Gaussian distribution, with means and variances derived from the coded $\hat{z}$. 
Therefore, the probability of the $i$th element is $p(\hat{y}_i|\hat{z}_i)=\mathcal{N}(\mu_i,\sigma_i^2)$ so that spatial redundancy is reduced for estimation. In the later channel-wise auto-regressive entropy model \cite{minnen2020channel}, $y$ is split along the channel dimension into $s$ slices $\{y_0, y_1, \dots, y_{s-1}\}$. The model utilizes both side information $\hat{z}$ and encoded channel slices $\hat{y}_{<i}=\{\hat{y}_0, \hat{y}_1, \dots, \hat{y}_{i-1}\}$ to encode the current channel slice $y_i$ with improved estimating performance. 
We formulate the entropy path as follows:
\begin{align}
  z = h_a(y;\phi_h), &\quad \hat{z} = Q(z), \quad (\mu_{side}, \sigma_{side}) = h_s(\hat{z}; \theta_h),\\
  (\mu_{i}, \sigma_{i}) &= e_i(\mu_{side}, \sigma_{side}, y_{<i}; \theta_{e_i}),\quad 0\leq i<s,
\end{align}
where $h_a$, $h_s$ and $e_i$ are parameterized by $\phi_h$, $\theta_h$ and $\theta_{e_i}$, respectively. 

LIC considers both rate and distortion, so the loss function is defined as:
\begin{equation}
\begin{split}
    \mathcal{L}&=\mathcal{R}(\hat{y})+\mathcal{R}(\hat{z})+\lambda \cdot \mathcal{D}(x,\hat{x}) \\
    &=\mathbb{E}[-\log_2(p_{\hat{y}|\hat{z}}(\hat{y}|\hat{z}))]+\mathbb{E}[-\log_2(p_{\hat{z}}(\hat{z}))]+\lambda \cdot \mathcal{D}(x,\hat{x}),
\end{split}
\label{eq:9}
\end{equation}
where $\mathbb{E}[-\log_2(p_{\hat{y}|\hat{z}}(\hat{y}|\hat{z}))]$ is the estimated rate of $\hat{y}$ given $\hat{z}$, and $\mathbb{E}[-\log_2(p_{\hat{z}}(\hat{z}))]$ is the estimated rate of $\hat{z}$. 
The trade-off between rate and distortion is controlled by a Lagrangian multiplier $\lambda$, where the rate is estimated by the entropy model and the distortion is measured by mean square error (MSE) during training. 
% We train distinct models for each $\lambda$ to achieve the final rate-distortion curve at various bit rates. 
In this work, we focus on modifications in $g_a$ and $g_s$.

\subsection{Method Overview}
\begin{figure}[tb]
  \centering
  \includegraphics[width=0.95\textwidth]{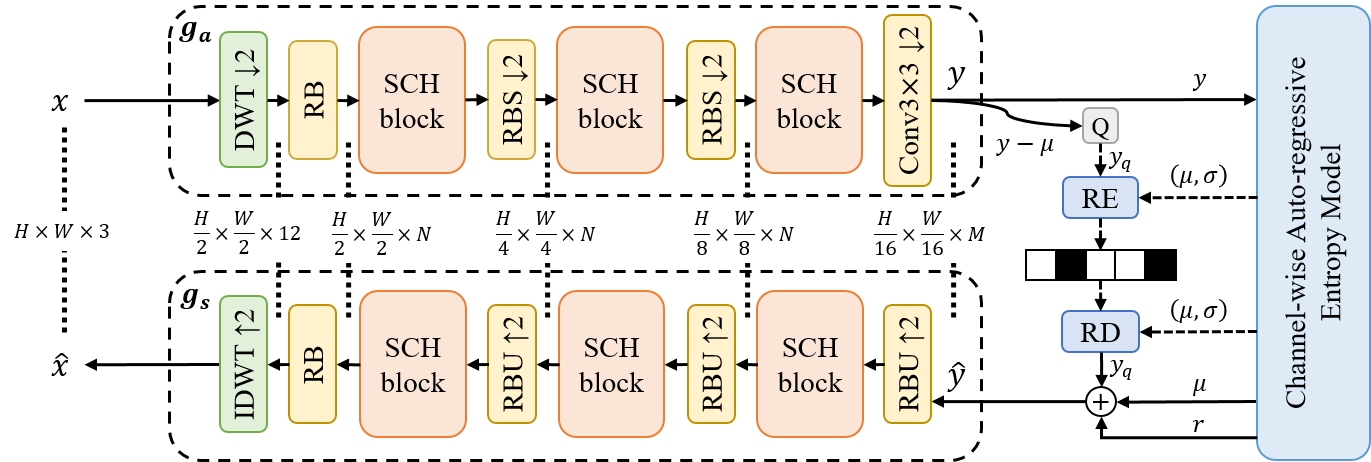}
  \caption{The overall architecture of our model. SCH is Space-Channel Hybrid block. DWT is the discrete wavelet transform and IDWT is the inverse transform. $\downarrow$ means down-sampling and $\uparrow$ means up-sampling. RB is Residual Block. RBS is Residual Block with Stride. RBU is Residual Block Up-sampling. RE is Range Encoder and RD is Range Decoder.}
  \label{fig:arch}
\end{figure}

The architecture of our SCH framework is illustrated in Fig.\ref{fig:arch}. 
We use the DWT module to decompose input image $x$ into four sub-images with different frequencies and realign them along the channel dimension. 
Inverse DWT projects four sub-images back to the 3-channel image.
Residual Block with Stride (RBS) and Residual Block Up-sampling (RBU) are proposed by \cite{cheng2020learned} for down-sampling and up-sampling. 
We place SCH blocks between the two successive residual blocks to learn spatial local information and channel-wise global information. 
The right part of Fig.\ref{fig:arch} contains the quantization, range encoder, channel-wise auto-regressive entropy model, and range decoder, following the design of \cite{liu2023learned}. 
Our framework features in SCH blocks with the proposed window-based channel attention module and DWT module. 
In the following sections, we explain the designs and properties of these modules in detail.

\subsection{Space-Channel Hybrid Block}
\begin{figure}[tb]
  \centering
  \begin{subfigure}{0.6\linewidth}
    \includegraphics[width=1.0\textwidth]{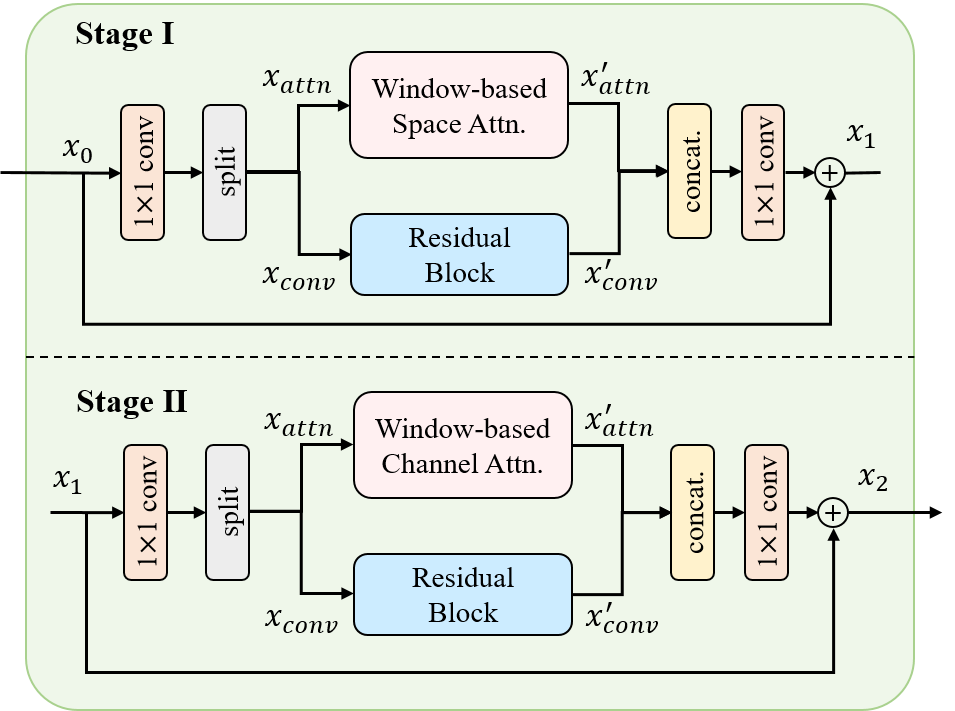}
    \caption{}
    \label{fig:SCAa}
  \end{subfigure}
  \hfill
  \begin{subfigure}{0.175\linewidth}
    \centering
    \includegraphics[width=0.725\textwidth]{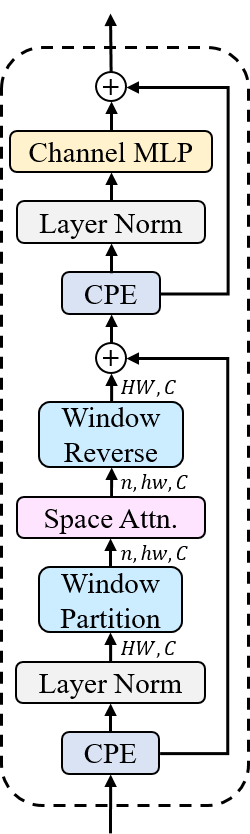}
    \caption{}
    \label{fig:SCAb}
  \end{subfigure}
  \begin{subfigure}{0.175\linewidth}
    \centering
    \includegraphics[width=0.725\textwidth]{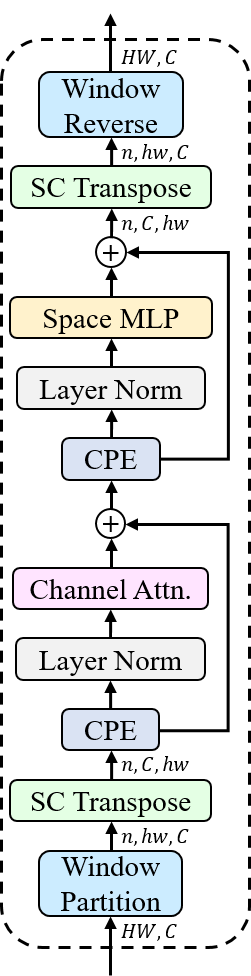}
    \caption{}
    \label{fig:SCAc}
  \end{subfigure}
  \caption{The proposed SCH block (left), window-based space attention module (middle), and window-based channel attention module (right). In (b) and (c), SC Transpose is Space-Channel Dimension Transposition. Modules with similar functions are marked in the same colors. We indicate tensor shapes before and after shape-transforming modules, where $n$, $h$, and $w$ are the number of windows, window height and width.}
  \label{fig:SCA}
\end{figure}

Taking inspiration from the TCM block \cite{liu2023learned}, we design our Space-Channel Hybrid block to efficiently capture both local and global dependencies in image features. 
We replace the learned relative positional encoding with Convolutional Positional Encoding (CPE) according to \cite{chu2022conditional} and replace the shifted-window attention from Swin-Transformer \cite{liu2021swin} with our window-based channel attention.

As shown in Fig.\ref{fig:SCAa}, in stage \uppercase\expandafter{\romannumeral1}, a $1 \times 1$ convolutional layer pre-processes the input tensor $x$, which is then evenly split into $x_{attn}$ and $x_{conv}$ along the channel dimension. 
This parallel processing design reduces the channel size in each branch, saving the computational cost of each processing module. 
Besides, each branch extracts features of diverse properties in parallel. 
Specifically, the window-based space attention module transforms $x_{attn}$ into $x'_{attn}$ to learn the transformer-based local spatial information, and the residual block transforms $x_{conv}$ into $x'_{conv}$ to learn the CNN-based local spatial information. 
Then, two tensors are concatenated in the channel dimension, and we fuse their information by the $1 \times 1$ convolutional layer. 
The skip connection between the input and the output facilitates gradient descent. 
In stage \uppercase\expandafter{\romannumeral2}, the window-based space attention module is replaced by the window-based channel attention module to capture global channel-wise dependencies in the tensor. 
All $1 \times 1$ convolutional layers in our design aggregate the information across different channels, and the number of feature channels remains unchanged. 
The above stages are formulated as follows:
\begin{align}
  x_{attn}, x_{conv} &= \text{Split}(\text{Conv1}\times\text{1}(x))\\
  x'_{attn}, x'_{conv} &= \text{Space}(x_{attn}) \text{ or }\text{Channel}(x_{attn}), \text{Res}(x_{conv})\\
  x_{out} &= x+\text{Conv1}\times\text{1}(\text{Concat}(x'_{attn}, x'_{conv})).
\end{align}
In summary, our SCH blocks capture local spatial dependencies via residual blocks and space attention modules, as well as global channel-wise dependencies via channel attention modules.

\subsection{Window-based Channel Attention Module}
The attention module in Vision Transformer (ViT) \cite{dosovitskiy2020image} defines the input shape as $L \times C$, where $L$ is the sequence length and $C$ is the channel size. 
To realize the Transformer-based channel attention, a straightforward attempt is to transpose two dimensions, which results in the tensor shape of $C \times L$, where $C$ becomes the sequence length and $L$ becomes the channel size. 
However, since image compression is a downstream task with various input sizes, $L$ is not a constant as the new channel size. 
As a result, we cannot design the linear projection and MLP (Multilayer Perceptron) layers if the channel size changes. 
Inspired by Swin-Transformer \cite{liu2021swin}, we partition the tensor into small windows so that the tensor shape becomes $n \times (window \; size)^2 \times C$, where $n$ is the number of windows and the sequence length $L$ equals to the fixed $(window\; size)^2$. 
With the dimension transpose and Convolutional Positional Encoding (CPE) from \cite{chu2022conditional}, we can formulate our window-based channel attention module as follows:
\begin{align}
  x' &= \text{CPE}_0(x) + \text{ChannelAttn}( \text{LayerNorm}(\text{CPE}_0(x)) )\\
  x_{out} &= \text{CPE}_1(x') + \text{MLP}( \text{LayerNorm}(\text{CPE}_1(x')) ).
\end{align}
The arrangement of the above modules follows \cite{ding2022davit}. 
As shown in Fig.\ref{fig:SCAc}, it should be noted that Window Partition and Window Reverse are rearranged so that CPE and MLP have the fixed channel size of $(window\; size)^2$. 
In addition, CPE encodes positional information in the channel dimension, and MLP fuses information along the space dimension after the channel attention, corresponding to the function of similar modules in Fig.\ref{fig:SCAb}.

% \subsubsection{Parameter Analysis}
% Our channel attention demonstrates a notable advantage in terms of parameter efficiency. 
% A conventional space attention module with a channel dimension of $C$ typically comprises four linear projection layers and one MLP layer, which has $4 \times (C \times C) + C \times 4C + 4C \times C = 12C^2$ parameters, assuming an MLP ratio of 4. 
% In contrast, our channel attention requires only $12(hw)^2$ parameters with the window size of $h \times w$. 
% Generally, $hw$ is smaller than $C$, so our channel attention has fewer parameters compared with the space attention. 
% Detailed derivation is reported in the supplementary materials.

\subsubsection{Advantages}
\begin{figure}[tb]
  \centering
  \begin{subfigure}{0.49\linewidth}
  \centering
    \includegraphics[width=0.95\textwidth]{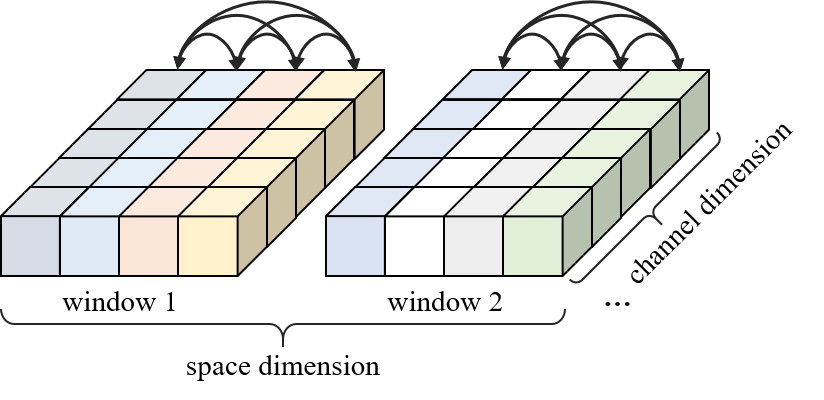}
    \caption{Window-based space attention}
    \label{fig:SCdemo_a}
  \end{subfigure}
  \hfill
  \begin{subfigure}{0.49\linewidth}
  \centering
    \includegraphics[width=0.95\textwidth]{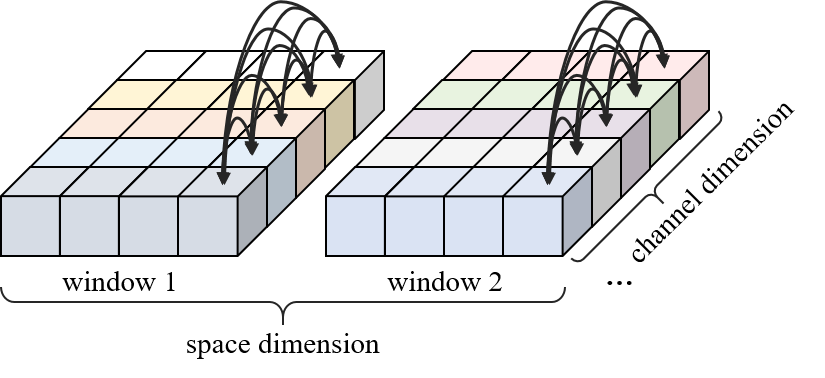}
    \caption{Window-based channel attention}
    \label{fig:SCdemo_b}
  \end{subfigure}
  \caption{Demonstration of window-based space attention and channel attention with window size $2 \times 2$ and channel size 5. In each window, (a) performs attention across space tokens and (b) performs attention across channel tokens. Different tokens are marked in different colors. The depth of the token is the actual channel size for computation.}
  \label{fig:SCdemo}
\end{figure}

Firstly, the proposed window-based channel attention captures global information with a large receptive field. 
Fig.\ref{fig:SCdemo} indicates that each channel token is global inside the window, providing the global view of the feature. 
Therefore, when we use channel tokens to calculate the channel attention map, formulated as $(C\times hw)\cdot(hw\times C)=C\times C$, it generates a global map shared throughout the window. 
With this map, attention computation can fuse multiple intact window slices globally, that is, $(C\times C)\cdot(C\times hw)=C\times hw$, and it outputs new global tokens with long-range information. 
To visualize the global effect of our window-based channel attention, we introduce the Effective Receptive Field (ERF) \cite{luo2016understanding}. 
It is derived by choosing a point in the feature map and doing the back-propagation since ERF can be obtained as gradients of one feature point to all pixels at the input. 
We select the feature from modules of the last block in $g_a$ to generate ERF. 
In Fig.\ref{fig:erf}, our window-based channel attention module has the largest ERF, and gradients focus on the texture and edges of petals and branches, indicating that the module can successfully capture global information. 
On the other hand, both residual block and space attention module have a smaller receptive field because of the local spatial information learning.
In addition, our window-based channel attention is influenced by wavelet transform, which assists the expansion of ERF as compared in Fig.\ref{fig:erf_c} and Fig.\ref{fig:erf_d}.

\begin{figure}[tb]
  \centering
  \begin{subfigure}{0.21\linewidth}
  \centering
    \includegraphics[width=2.56cm]{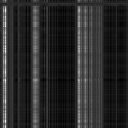}
    \caption{}
    \label{fig:ca0}
  \end{subfigure}
  \hfill
  \begin{subfigure}{0.21\linewidth}
  \centering
    \includegraphics[width=2.56cm]{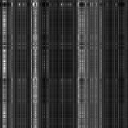}
    \caption{}
    \label{fig:ca1}
  \end{subfigure}
  \hfill
  \begin{subfigure}{0.21\linewidth}
  \centering
    \includegraphics[width=2.56cm]{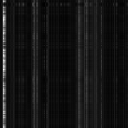}
    \caption{}
    \label{fig:ca2}
  \end{subfigure}
  \hfill
  \begin{subfigure}{0.075\linewidth}
  \centering
    \includegraphics[width=0.32cm]{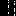}
    \caption{}
    \label{fig:ca3}
  \end{subfigure}
  \begin{subfigure}{0.075\linewidth}
  \centering
    \includegraphics[width=0.32cm]{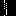}
    \caption{}
    \label{fig:ca4}
  \end{subfigure}
  \begin{subfigure}{0.075\linewidth}
  \centering
    \includegraphics[width=0.32cm]{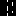}
    \caption{}
    \label{fig:ca5}
  \end{subfigure}
  \caption{Channel attention maps on \emph{kodim07} from our module and DaViT \cite{ding2022davit}. Our window-based channel attention offers $n$ windows $H$ heads $C\times C$ maps, and we randomly select three maps \ref{fig:ca0}, \ref{fig:ca1} and \ref{fig:ca2} from different windows of the first head. DaViT offers $H$ heads $C_g\times C_g$ maps, where $C=H\times C_g$, and we visualize three maps \ref{fig:ca3}, \ref{fig:ca4} and \ref{fig:ca5} from three heads. $n$, $H$, and $C$ are 96, 8, and 128, respectively}
  \label{fig:attn_map}
\end{figure}

Secondly, our window-based channel attention captures more global information than existing channel attention methods. 
SENet \cite{hu2018squeeze} squeezes global spatial information through global average pooling to obtain $C \times 1$ channel dependencies.
CBAM \cite{woo2018cbam} includes max-pooling into squeezing operation to obtain two different $C \times 1$ channel dependencies.
DaViT \cite{ding2022davit} introduces channel group attention and provides $N_g$ groups $C_g \times C_g$ attention maps, where $C = N_g \times C_g$.
Supposing our window-based channel attention divides the feature map into $n$ windows, it can generate $n$ different $C \times C$ channel attention maps, which contain more information than aforementioned methods.
As shown in Fig.\ref{fig:attn_map}, our window-based channel attention provides larger and more maps than modules from DaViT.
Moreover, we can see the variety among these maps, further proving the effectiveness of our channel attention design. 
Although channel attention maps from DaViT or other methods have similar diversities, they contain significantly less channel information than ours.
We suggest preserving more diverse channel information assists global information learning, and we will compare the compression performance of channel attention designs in Section \ref{sec:ablation}.

\subsection{Wavelet Transform Module}
\begin{figure}[tb]
  \begin{subfigure}{0.24\linewidth}
      \centering
      \includegraphics[width=1\textwidth]{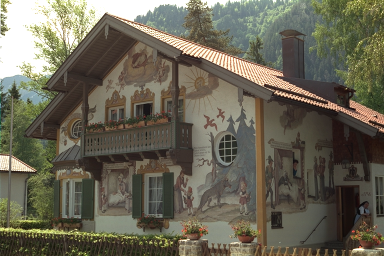}
      \caption{$I_{LL}$}
  \end{subfigure}
  \hfill
  \begin{subfigure}{0.24\linewidth}
      \centering
      \includegraphics[width=1\textwidth]{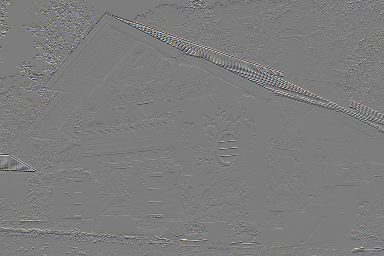}
      \caption{$I_{LH}$}
  \end{subfigure}
  \hfill
  \begin{subfigure}{0.24\linewidth}
      \centering
      \includegraphics[width=1\textwidth]{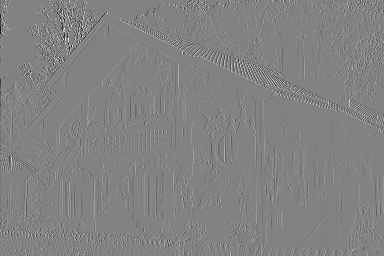}
      \caption{$I_{HL}$}
  \end{subfigure}
  \hfill
  \begin{subfigure}{0.24\linewidth}
      \centering
      \includegraphics[width=1\textwidth]{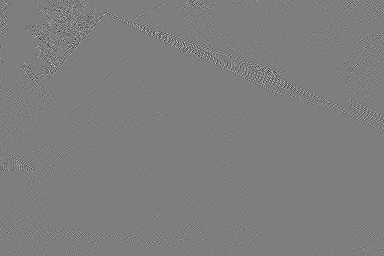}
      \caption{$I_{HH}$}
  \end{subfigure}
  \caption{Visualization of Haar discrete wavelet transform on \emph{kodim24}.}
  \label{fig:visual_haar}
\end{figure}

To further enhance receptive fields, we introduce a wavelet transform module to process the input image. 
We choose the Haar wavelet because of its simplicity and efficiency \cite{liu2018multi,liu2019multi,wang2020multi,jeevan2022wavemix,yao2022wave,li2024ewt}, whose four filters are defined as,
\begin{equation}
    f_{LL} = \frac{1}{2}\begin{bmatrix}
    1 & 1 \\
    1 & 1
    \end{bmatrix}\text{,}\quad
    f_{LH} = \frac{1}{2}\begin{bmatrix}
    1 & -1 \\
    1 & -1
    \end{bmatrix}\text{,}\quad
    f_{HL} = \frac{1}{2}\begin{bmatrix}
    1 & 1 \\
    -1 & -1
    \end{bmatrix}\text{,}\quad
    f_{HH} = \frac{1}{2}\begin{bmatrix}
    1 & -1 \\
    -1 & 1
    \end{bmatrix}\text{.}
  \label{eq:important}
\end{equation}
We can see that $f_{LL} \otimes x$ performs a sum-pooling operation that outputs the down-sampled low-frequency approximation of the original image $x$, while the other filters aim to obtain high-frequency information. 
As shown in Fig.\ref{fig:visual_haar}, $I_{LL}$ is the down-sampled feature of the original image. 
$I_{LH}$ and $I_{HL}$ are the horizontal details and vertical details of the image, describing the edge features of two directions. 
$I_{HH}$ is the diagonal details of the image. 
Since wavelet transform decomposes the image into components of different frequencies, it guides the neural network in learning frequency correlations and assists in learning more complicated textures.

After that, we stack four sub-images in the channel dimension and use a residual block to project images from RGB space to latent space. 
Therefore, the wavelet transform acts as a role of frequency-dependent down-sampling to enlarge receptive fields of the model. It affects feature capturing of residual blocks, space attention modules, and channel attention modules, but it mainly assists channel attention for global information learning because of large receptive fields. Meanwhile, the wavelet transform is effective because it contains no parameters for learning.

% Secondly, our channel attention has a lower computational complexity. 
% For one space attention block with the input size $L \times C$, the computational cost of four linear projection layers is $4 \times (2 \times L \times C \times C) = 8LC^2$. 
% The cost of attention is $2 \times L \times C \times L + 2 \times L \times L \times C = 4L^2C$. 
% The cost of MLP is $2 \times L \times C \times 4C + 2 \times L \times 4C \times C = 16LC^2$. 
% The total computational cost of the space attention is $8LC^2+4L^2C+16LC^2=24LC^2+4L^2C$, while the cost of the channel attention is $24CL^2+4C^2L$. 
% Since $L=h \times w < C$, the proposed channel attention is more efficient. 

% , assisting the global information learning for our channel attention.
% \begin{figure}[tb]
%   \centering
%   \includegraphics[width=0.6\textwidth]{fig/5.png}
%   \caption{Procedure of the sub-images realignment and the Residual Block projection.}
%   \label{fig:waveres}
% \end{figure}

\section{Experiments}
\subsection{Experimental Setup}
We implement the proposed model on CompressAI \cite{begaint2020compressai} platform. 
For training, we choose the largest 300k images from the ImageNet training set \cite{deng2009imagenet} and randomly crop them into the size of $256 \times 256$. 
We use the Adam optimizer \cite{adam2015} with a batch size of 8. 
The learning rate is initialized as $1 \times 10^{-4}$ and scheduled by the PyTorch \cite{paszke2019pytorch} learning rate scheduler ReduceLROnPlateau with the patience of 5 epochs and the factor of 0.3. 
The model is optimized by the loss in Eq.\ref{eq:9}, where the distortion is measured in mean square error (MSE). 
We set the Lagrangian multiplier $\lambda$ as \{0.0025, 0.0035, 0.0067, 0.013, 0.025, 0.05\}. 
The models are trained for 3.5M iterations on average.

For our architecture in Fig.\ref{fig:arch}, $N$ is 256, $M$ is 320, and the channel size of $z$ is 192. 
We stack \{2,4,2\} blocks in $g_a$ and $g_s$, respectively, and the window size is $8 \times 8$ in each block to make it comparable to other methods about the parameter quantity. 
We use NVIDIA A100 to train and evaluate our model.

\subsection{Performance}
\subsubsection{Rate-Distortion Performance}
We evaluate our model on four commonly used datasets, including Kodak image set \cite{Kodak1992}, Tecnick test set \cite{asuni2014testimages}, CLIC professional validation set \cite{toderici2020workshop}, and CLIC 2021 test set \cite{toderici2021workshop}. 
PSNR is used to measure the distortion, while bits per pixel (bpp) are used to evaluate bit rates.

\begin{figure}[tb]
  \centering
  \begin{subfigure}{0.495\linewidth}
  \centering
    \includegraphics[width=1\textwidth]{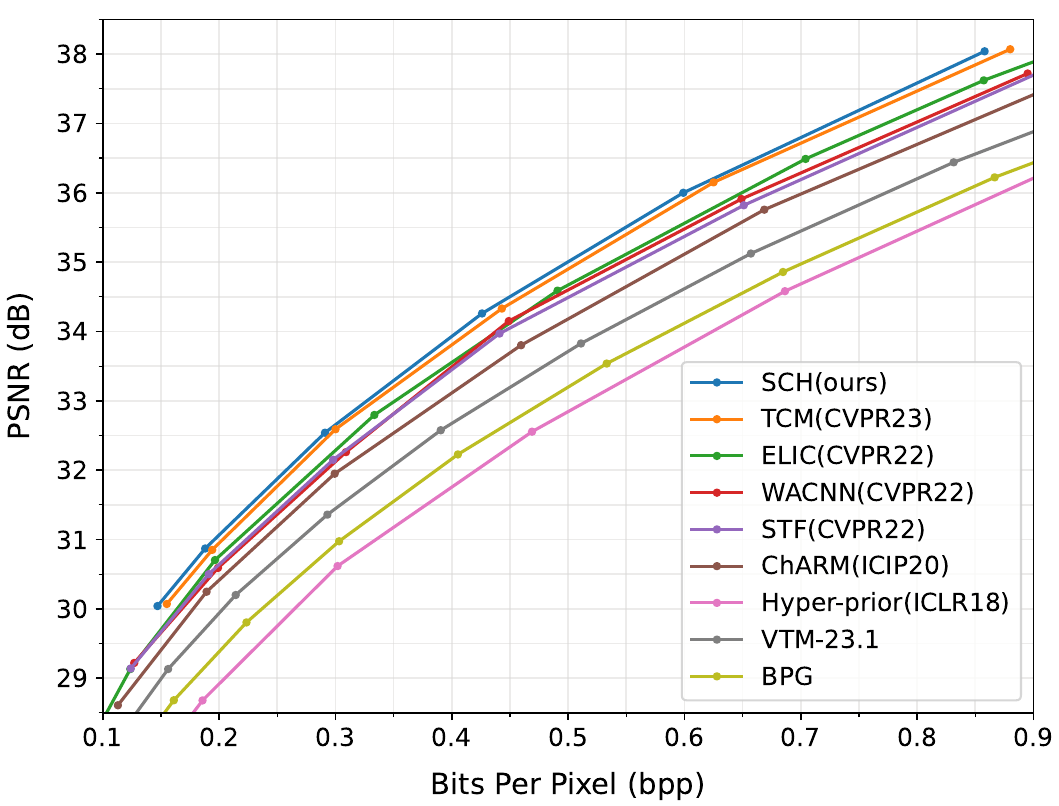}
    \caption{Kodak ($768\times512$)}
    \label{fig:psnr_a}
  \end{subfigure}
  \hfill
  \begin{subfigure}{0.495\linewidth}
  \centering
    \includegraphics[width=1\textwidth]{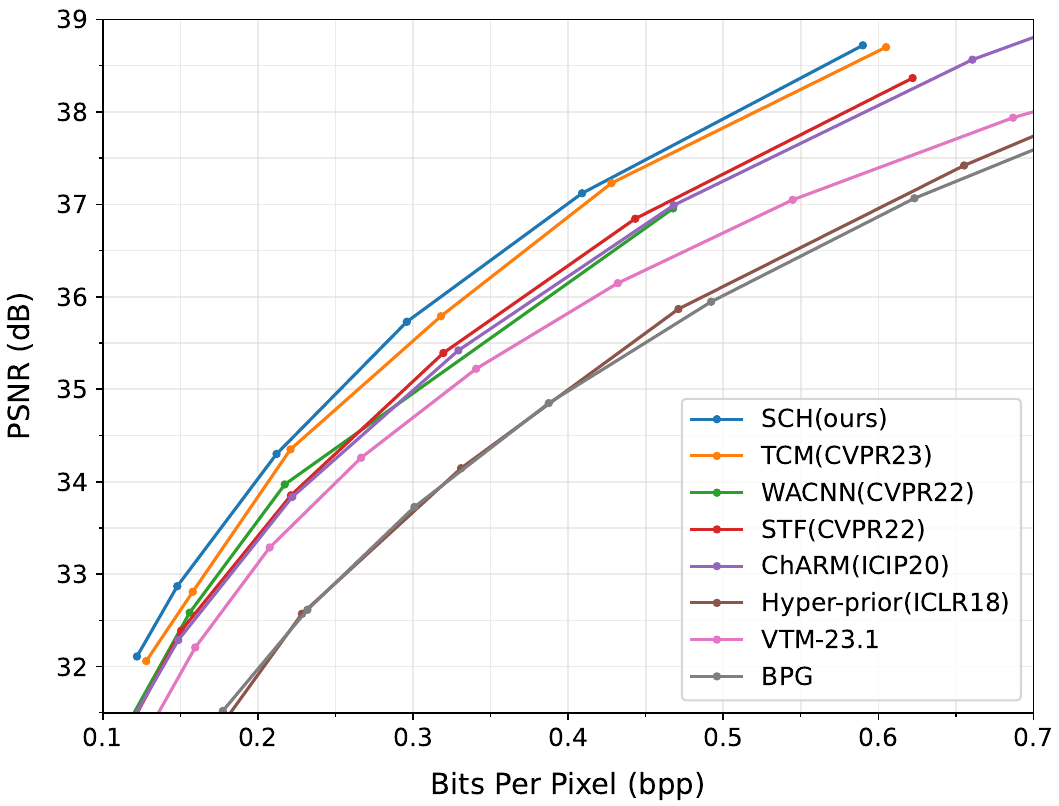}
    \caption{Tecnick ($1200\times1200$)}
    \label{fig:psnr_b}
  \end{subfigure}\\
  \begin{subfigure}{0.495\linewidth}
  \centering
    \includegraphics[width=1\textwidth]{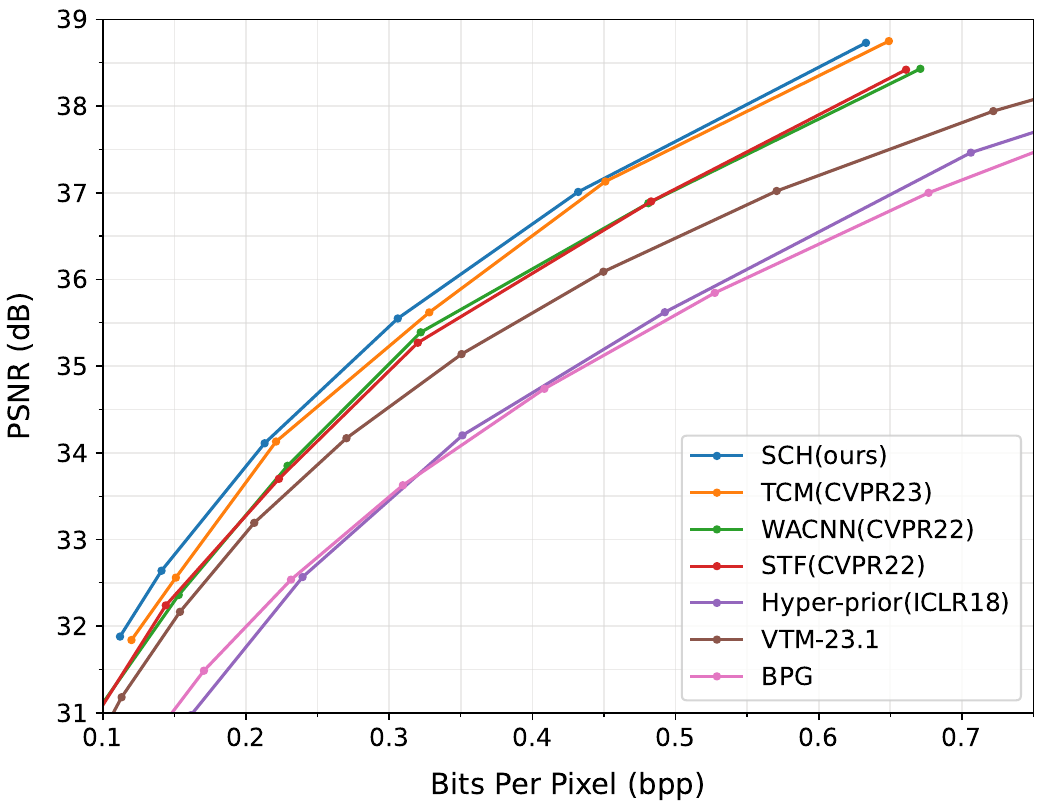}
    \caption{CLIC Professional Valid (up to $2048\times1370$)}
    \label{fig:psnr_c}
  \end{subfigure}
  \hfill
  \begin{subfigure}{0.495\linewidth}
  \centering
    \includegraphics[width=1\textwidth]{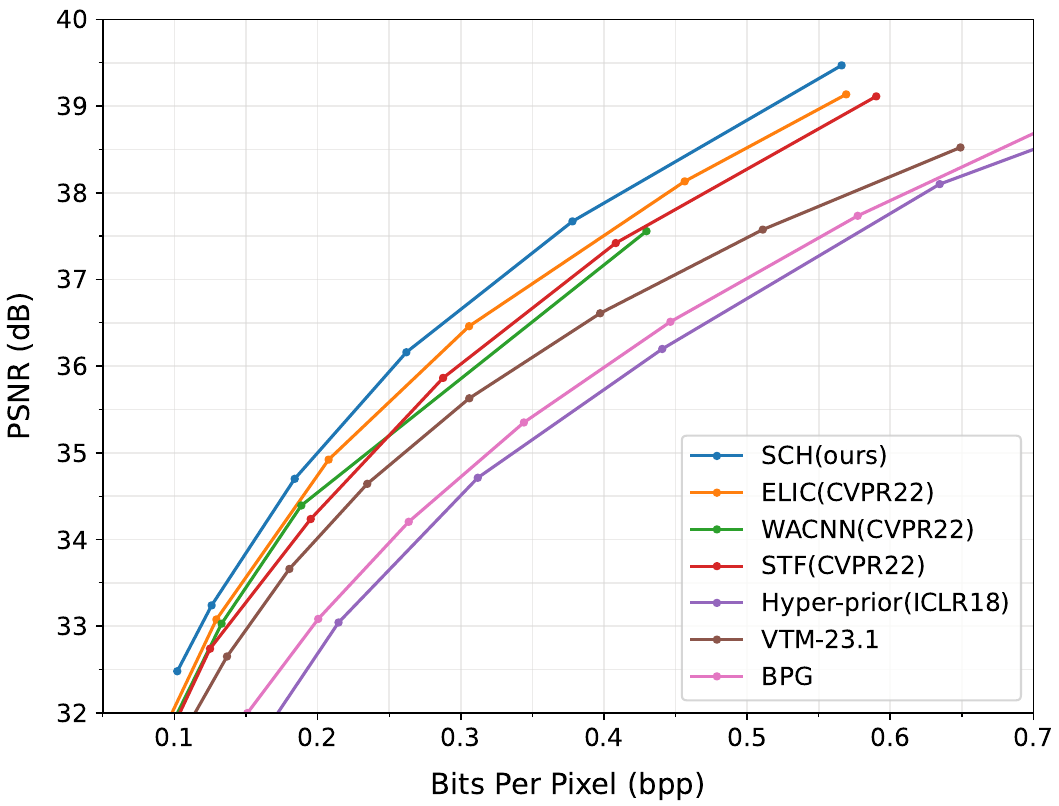}
    \caption{CLIC Test 2021 (up to $2048\times1415$)}
    \label{fig:psnr_d}
  \end{subfigure}
  \caption{RD performance evaluation on four standard datasets with various resolutions.}
  \label{fig:psnr}
\end{figure}

The Rate-Distortion (RD) performance on the Kodak dataset \cite{Kodak1992} is illustrated in Fig. \ref{fig:psnr_a}. 
We compare our SCH with recent models \cite{liu2023learned,he2022elic,zou2022devil}, classic models \cite{minnen2020channel,ballé2018variational}, BPG \cite{bellard2014bpg} and VTM-23.1 of Versatile Video Coding (VVC) \cite{wien2020versatile}. 
Our approach outperforms the state-of-the-art (SOTA) method \cite{liu2023learned} by 0.125dB on average.
Additionally, results on the Tecnick \cite{asuni2014testimages}, CLIC professional validation set \cite{toderici2020workshop}, and CLIC 2021 test set \cite{toderici2021workshop} are depicted in Fig. \ref{fig:psnr_b}, \ref{fig:psnr_c}, and \ref{fig:psnr_d}, respectively. 
We achieve up to 0.31dB improvements across three datasets, suggesting the robustness of our method and its ability to achieve SOTA performances for diverse resolutions. 
The proposed channel attention module excels at capturing global dependencies in images. Therefore, our SCH achieves more performance gain on later datasets with larger images.

To quantify the performance of our method, we present the BD-rate \cite{bjontegaard2001calculation} computed from the RD data as the metric. 
We set the results of the latest VTM-23.1 as anchor rates (BD-rate = 0\%) in four datasets. 
Our method reduces BD-rate by 18.54\%, 23.98\%, 22.33\%, and 24.71\% on the Kodak \cite{Kodak1992}, Tecnick \cite{asuni2014testimages}, CLIC professional validation \cite{toderici2020workshop}, and CLIC 2021 test \cite{toderici2021workshop} datasets, as presented in Table \ref{bdrate}. 
Compared with TCM \cite{liu2023learned}, our method achieves a solid performance gain on BD-rate, ranging from 1.91\% to 3.46\% across four datasets.

\begin{table}[tb]
  \caption{
    BD-Rate (\%) comparison in four datasets. "-" means the author did not provide models or results. The best rate is shown in bold. The anchor is VTM-23.1 \cite{wien2020versatile}.
  }
  \label{bdrate}
  \tabcolsep=7.4pt
  \centering
  \begin{tabular}{@{}lcccc@{}}
    \toprule
    Method & Kodak & Tecnick & CLIC Pro Val & CLIC'21 Test \\
    \midrule
    VTM-23.1 \cite{wien2020versatile}                & 0 & 0 & 0 & 0\\
    Hyper-prior (ICLR18) \cite{ballé2018variational} & 30.32 & 25.19 & 29.76 & 30.59\\
    ChARM (ICIP20) \cite{minnen2020channel}          & -4.06 & -10.05 & - & -\\
    STF (CVPR22) \cite{zou2022devil}                 & -9.05 & -10.98 & -10.74 & -11.74\\
    WACNN (CVPR22)  \cite{zou2022devil}              & -9.51 & -11.97 & -11.09 & -14.07\\
    ELIC (CVPR22)  \cite{he2022elic}                 & -12.13 & - & - & -18.35\\
    TCM (CVPR23) \cite{liu2023learned}               & -16.63 & -20.52 & -18.89 & -\\
    \textbf{SCH} (ours) & \textbf{-18.54} & \textbf{-23.98} & \textbf{-22.33} & \textbf{-24.71}\\
  \bottomrule
  \end{tabular}
\end{table}

\subsubsection{Qualitative Results}
\begin{figure}[tb]
  \centering
  \begin{subfigure}{0.2425\linewidth}
  \centering
    \includegraphics[width=1\textwidth]{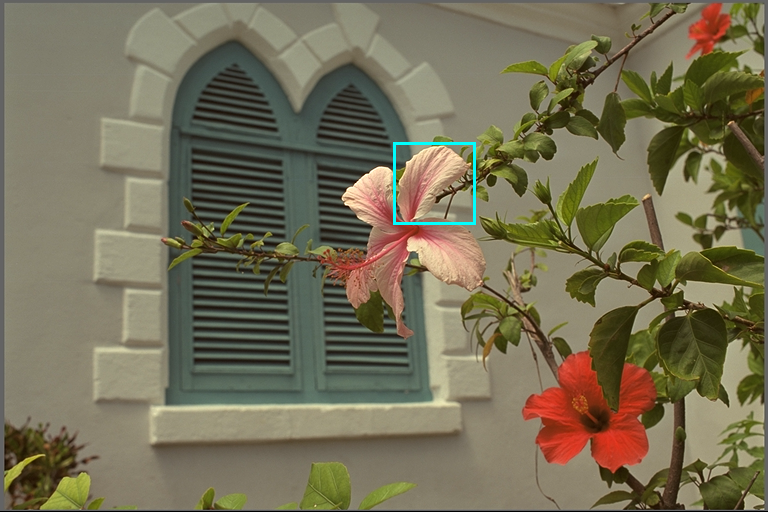}
    \caption{Ground Truth}
    \label{fig:visual_a}
  \end{subfigure}
  \begin{subfigure}{0.2425\linewidth}
  \centering
    \includegraphics[width=1\textwidth]{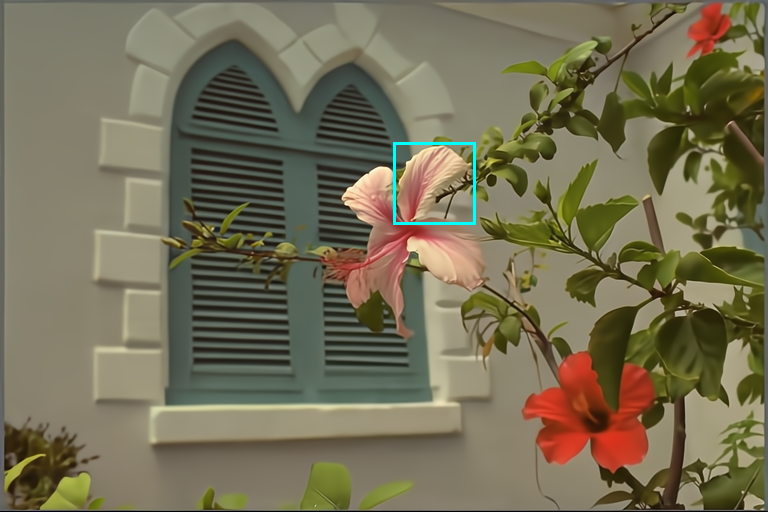}
    \caption{SCH (ours)}
    \label{fig:visual_b}
  \end{subfigure}
  \begin{subfigure}{0.2425\linewidth}
  \centering
    \includegraphics[width=1\textwidth]{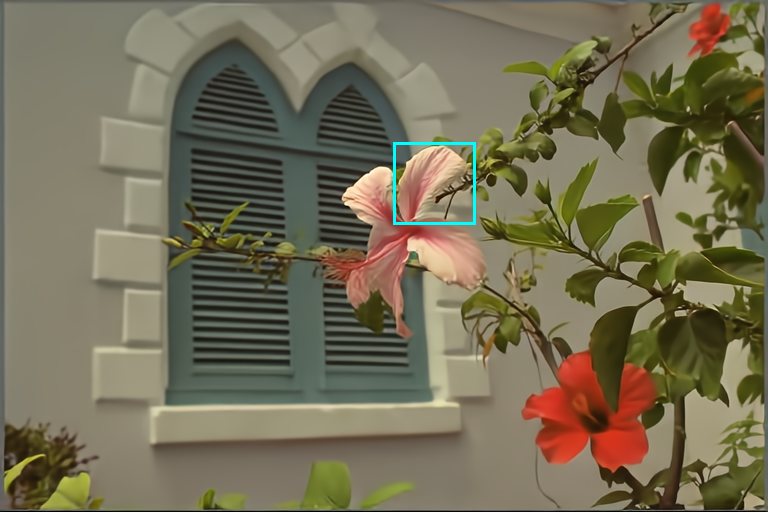}
    \caption{STF}
    \label{fig:visual_c}
  \end{subfigure}
  \begin{subfigure}{0.2425\linewidth}
  \centering
    \includegraphics[width=1\textwidth]{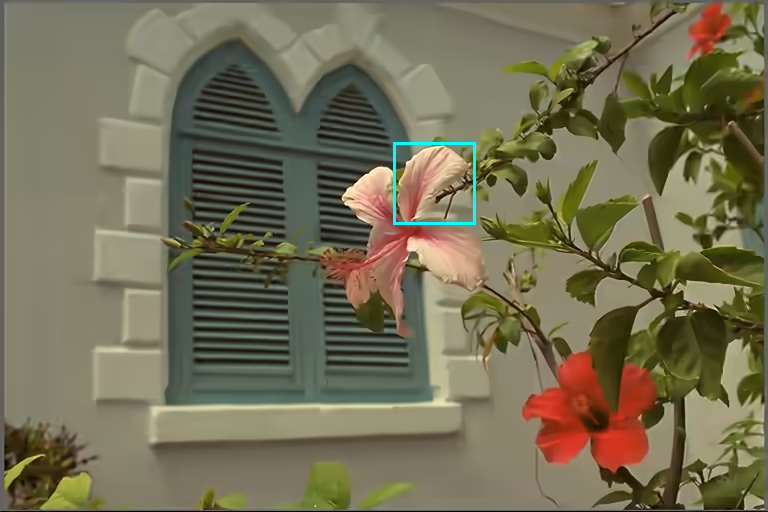}
    \caption{VTM-23.1}
    \label{fig:visual_d}
  \end{subfigure}\\
  
  \begin{subfigure}{0.2425\linewidth}
  \centering
    \includegraphics[width=0.6\textwidth]{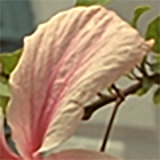}
    \text{bpp/PSNR}
    \label{fig:visual_e}
  \end{subfigure}
  \begin{subfigure}{0.2425\linewidth}
  \centering
    \includegraphics[width=0.6\textwidth]{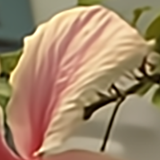}
    \text{0.146/33.159dB}
    \label{fig:visual_f}
  \end{subfigure}
  \begin{subfigure}{0.2425\linewidth}
  \centering
    \includegraphics[width=0.6\textwidth]{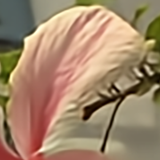}
    \text{0.144/32.617dB}
    \label{fig:visual_g}
  \end{subfigure}
  \begin{subfigure}{0.2425\linewidth}
  \centering
    \includegraphics[width=0.6\textwidth]{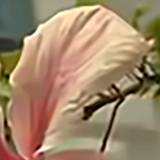}
    \text{0.157/31.963dB}
    \label{fig:visual_h}
  \end{subfigure}\\
  \caption{Visualization of the reconstructed \emph{kodim07} images from Kodak. Our SCH is compared with an available learned method STF \cite{zou2022devil} and the latest traditional codec VTM-23.1 \cite{wien2020versatile}. The metrics are [bpp$\downarrow$/PSNR$\uparrow$]. The second row contains the enlarged portions from the cyan rectangles of the first row, respectively.}
  \label{fig:visual}
\end{figure}

Fig.\ref{fig:visual} visualizes the reconstructed images (\emph{kodim07}) from our SCH, an available learned model STF \cite{zou2022devil} and the latest traditional codec VTM-23.1 \cite{wien2020versatile}. 
We choose the same image as Fig.\ref{fig:erf} and enlarge the same area to compare the texture and edges reconstructed by different methods.
Firstly, our method offers the clearest texture details of the petal with a low bit rate, which benefits from the global information learning provided by our window-based channel attention.
Secondly, our method reconstructs the sharpest and most accurate edges of the petal.
It indicates that our SCH has a strong capability of distinguishing complicated foregrounds and backgrounds, which benefits from the combination of local space modules and global channel modules. 
We provide more qualitative results in the supplementary materials.

\subsubsection{Codec Efficiency Analysis}
We compare our SCH with available recent models \cite{zou2022devil,liu2023learned} and VTM-23.1 \cite{wien2020versatile} on the efficiency. 
For the encoding time and decoding time, our method is comparable with TCM \cite{liu2023learned}. 
When considering parameters and BD-rate, our method achieves the SOTA BD-rate while maintaining the second lowest quantity of parameters. Detailed parameter analysis is reported in the supplementary materials.

\begin{table}[tb]
  \caption{
    Efficiency comparison on Kodak dataset, including average encoding time, average decoding time, model parameter,rs and BD-rate (\%).
  }
  \label{efficiency}
  \tabcolsep=7.4pt
  \centering
  \begin{tabular}{@{}lcccc@{}}
    \toprule
    Method & Enc (s) & Dec (s) & Parameters (M) & BD-rate$\downarrow$\\
    \midrule
    VTM-23.1 \cite{wien2020versatile} & 108.70 & 0.173 & - & 0\\
    STF (CVPR22) \cite{zou2022devil} & 0.123 & 0.151 & 99.86 & -9.05\\
    WACNN (CVPR22) \cite{zou2022devil} & 0.102 & 0.133 & 75.24 & -9.51\\
    TCM (CVPR23) \cite{liu2023learned}  & 0.193 & 0.172 & 76.57 & -16.63\\
    SCH (ours) & 0.184 & 0.181 & 75.79 & -17.77\\
  \bottomrule
  \end{tabular}
\end{table}

\subsection{Ablation Studies}
\label{sec:ablation}
\begin{figure}[tb]
  \centering
  \begin{subfigure}{0.495\linewidth}
  \centering
    \includegraphics[width=1\textwidth]{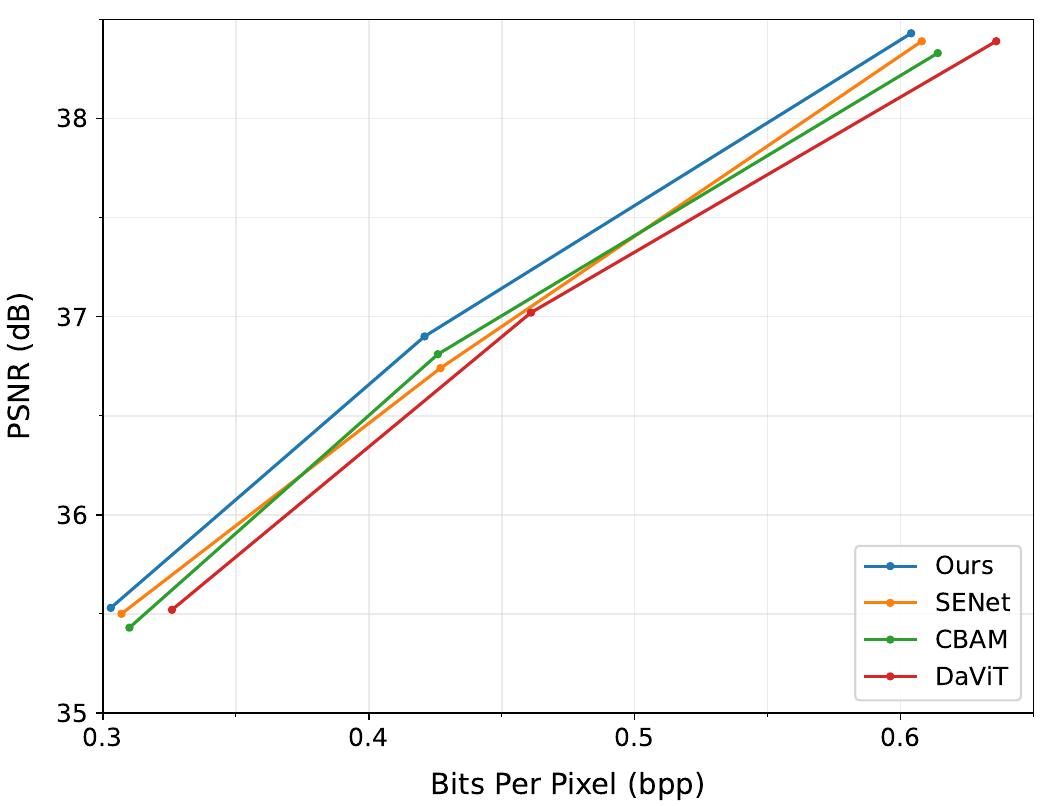}
    \caption{Channel Attention Module}
    \label{fig:ablation_a}
  \end{subfigure}
  \hfill
  \begin{subfigure}{0.495\linewidth}
  \centering
    \includegraphics[width=1\textwidth]{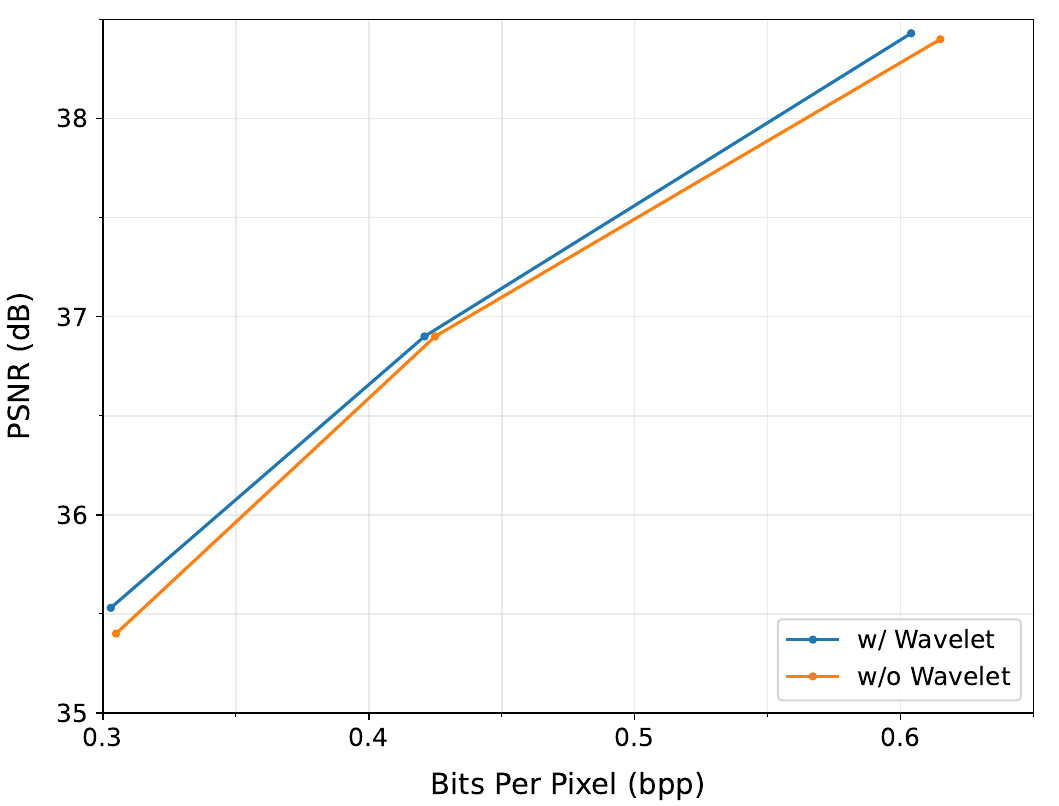}
    \caption{Wavelet Transform Module}
    \label{fig:ablation_b}
  \end{subfigure}\\
  \caption{Ablation studies on Tecnick dataset. (a) Channel attention modules (Ours, SENet \cite{hu2018squeeze}, CBAM \cite{woo2018cbam}, and DaViT \cite{ding2022davit}). (b) Wavelet transform module.}
  \label{fig:ablation}
\end{figure}

\subsubsection{Channel Attention Module}
We conducted an ablation study as depicted in Fig.\ref{fig:ablation_a} to validate the advantage of our window-based channel attention module. Each model is trained for 1.5M iterations on ImageNet \cite{deng2009imagenet}.
We compared our approach against existing channel attention modules from SENet \cite{hu2018squeeze}, CBAM \cite{woo2018cbam}, and DaViT \cite{ding2022davit}. 
The result indicates that our approach is better than CNN-based channel attention modules, and the Transformer-based channel attention module from DaViT is unsuitable for image compression, proving the superiority of our window-based channel attention module.

\subsubsection{Wavelet Transform Module}
In Fig.\ref{fig:ablation_b}, we compare the cases about using the wavelet transform module or not. 
It can be observed that our wavelet transform module brings a solid gain in RD performance.
The results indicate that enlarging receptive fields is beneficial to the performance enhancement of LIC.

\section{Conclusion}
In this paper, we propose a novel Space-Channel Hybrid (SCH) framework, which contains residual blocks and space attention modules for local information learning, as well as channel attention modules for global information learning. 
Our window-based channel attention module is the first to include a window partition, which captures more global information than other solutions and significantly enlarges receptive fields for RD performance gain in LIC.  
To further enhance receptive fields, we integrate a Haar DWT module into our framework to process raw images. 
Extensive experimental results demonstrate SOTA performances of our approach across four datasets with various resolutions.

Although our method is comparable with existing LIC works in terms of computational cost, it is still too high for applications on mobile devices. 
We expect that this problem can be addressed by model compression techniques, such as pruning, quantization, knowledge distillation, and structural re-parameterization.

\subsubsection{\ackname}
This work is supported by National Natural Science Foundation of China (No. 62331014) and Center of Computational Science and Engineering at Southern University of Science and Technology.

% \clearpage  % TODO REVIEW/FINAL: This \clearpage needs to be removed from both review and camera-ready versions.

\bibliographystyle{splncs04}
\bibliography{main}
\end{document}